\documentclass[10pt,journal,comsoc]{IEEEtran}
\usepackage{cite}
\usepackage{bm}
\usepackage{amsmath}
\usepackage{extarrows}
\usepackage{amssymb}
\usepackage{graphicx}
\usepackage{color}
\usepackage{enumerate}
\usepackage{bookmark} 
\graphicspath{{figure}}
\usepackage{setspace}
\usepackage{subfigure}
\usepackage{algorithm}
\usepackage{algorithmicx}
\usepackage{multirow}
\usepackage{stfloats}

\newcommand{\mr}{\mathrm} 

\newcommand{\BE}{\begin{equation}}
\newcommand{\EE}{\end{equation}}
\newcommand{\BS}{\begin{subequations}}
	\newcommand{\ES}{\end{subequations}}
\renewcommand{\bf}{\bm}

\newtheorem{theorem}{Theorem}

\newtheorem{assumption}{Assumption}

\newtheorem{property}{Property}

\newtheorem{lemma}{Lemma}

\allowdisplaybreaks \allowdisplaybreaks[2]

\ifCLASSINFOpdf
\graphicspath{{figure/}}

\else

\fi

\begin{document}

\title{{Capacity Optimal Coded Generalized MU-MIMO}}

\author{\IEEEauthorblockN{Yuhao Chi\IEEEauthorrefmark{1}, \emph{Member, IEEE}, Lei Liu\IEEEauthorrefmark{2}, \emph{Member, IEEE}, Guanghui Song\IEEEauthorrefmark{1}, \emph{Member, IEEE}, \\ Ying Li\IEEEauthorrefmark{1}, \emph{Member, IEEE}, Yong Liang Guan\IEEEauthorrefmark{3}, \emph{Senior Member, IEEE}, and Chau Yuen\IEEEauthorrefmark{4}, \emph{Fellow, IEEE}}\\
{\normalsize
	\IEEEauthorrefmark{1}State Key Lab of ISN, Xidian University, China,
	\IEEEauthorrefmark{2}Japan Advanced Institute of Science and Technology (JAIST), Japan\\
	\IEEEauthorrefmark{3}Nanyang Technological University, Singapore,
	\IEEEauthorrefmark{4}Singapore University of Technology and Design, Singapore
}
}

\maketitle
\pagestyle{empty}
\thispagestyle{empty}
\vspace{-1cm}
\begin{abstract}
	With the complication of future communication scenarios, most conventional signal processing technologies of multi-user multiple-input multiple-output (MU-MIMO) become unreliable, which are designed based on ideal assumptions, such as Gaussian signaling and independent identically distributed (IID) channel matrices. As a result,  this paper considers a generalized MU-MIMO (GMU-MIMO) system with more general assumptions, i.e., arbitrarily fixed input distributions, and general unitarily-invariant channel matrices. However, there is still no accurate capacity analysis and capacity optimal transceiver with practical complexity for GMU-MIMO under the constraint of coding. To address these issues, inspired by the replica method, the constrained sum capacity of coded GMU-MIMO with fixed input distribution is calculated by using the celebrated mutual information and minimum mean-square error (MMSE) lemma and the MMSE optimality of orthogonal/vector approximate message passing (OAMP/VAMP). Then, a capacity optimal multi-user OAMP/VAMP receiver is proposed, whose achievable rate is proved to be equal to the constrained sum capacity. Moreover, a design principle of multi-user codes is presented for the multi-user OAMP/VAMP, based on which a kind of practical multi-user low-density parity-check (MU-LDPC) code is designed. Numerical results show that finite-length performances of the proposed MU-LDPC codes with multi-user OAMP/VAMP are about 2~dB away from the constrained sum capacity and outperform those of the existing state-of-art methods.
\end{abstract}

\vspace{-0.3cm}
\section{Introduction}
To support massive connections of the fast-growing Internet of Things and wireless communications~\cite{Ding2021IeeeCM}, multi-user multiple-input multiple-output (MU-MIMO)\cite{Rusek2013,Miao2013,GoutayJSAC2021} was employed. Note that well-designed signal processing technologies of MU-MIMO are often studied based on ideal assumptions, which are Gaussian signaling and independent identically distributed (IID) channels, such as Rayleigh fading matrices. Nevertheless, with the diversification of future communication services~\cite{Ding2021IeeeCM}, these ideal assumptions are not still sufficient to accurately represent complex communication scenarios. 
\vspace{-0.2cm}
\subsection{System Assumptions of Generalized MU-MIMO}
This paper considers an uplink generalized MU-MIMO (GMU-MIMO) with the more general assumptions: 1) arbitrarily fixed input distributions, 2) general unitarily-invariant channel matrices, 3) massive users and antennas, 4) channel state information (CSI) only available at the receiver. To illustrate the rationality and practicality of the assumptions, we explain the necessity of these assumptions. 

\begin{enumerate}
\item Arbitrarily fixed input distributions: practical transmitters employ non-Gaussian discrete signalings, such as quadrature phase-shift keying (QPSK) and quadrature amplitude modulation (QAM). Meanwhile, the transmitted signalings are fixed during transmission.
\item General unitarily-invariant channel matrices: most practical channel matrices are not strictly IID. To cope with a class of non-IID channels, general unitarily-invariant channel matrices~\cite{MaAcess2017, MaTWC2019, Poor2021TWC} are considered, covering a variety of fading models including Rayleigh fading matrices, certain ill-conditioned and correlated matrices.
\item CSI only available at the receiver: in the uplink MU-MIMO, it is often infeasible to acquire the exact CSI for users. As a result, CSI is assumed to be available at the receiver~\cite{MaAcess2017, MaTWC2019, Poor2021TWC,LeiTSP2019,YuhaoTWC2018}. 
\end{enumerate}

However, under the constraints of channel coding, the information-theoretic limit and the low-complexity information-theoretically optimal receiver of GMU-MIMO are still open issues. 

\subsection{Information Theoretical Limit of MU-MIMO}

The information-theoretic limit of MU-MIMO is channel capacity. It is well known that capacity is defined by default as the maximum mutual information over all possible choices of the input distribution. However, due to the constraint of the arbitrarily fixed input distribution, we employ a constrained capacity in GMU-MIMO, which is defined as the mutual information under a fixed input distribution $x \sim P_X(x)$. To avoid confusing usage, capacity, named ``Gaussian capacity'', is achieved by Gaussian signaling. The constrained capacity is employed when given input distribution $x \sim P_X(x)$.

For available CSI at the transceiver, a kind of iterative water-filling method was proposed to obtain the Gaussian sum capacity of MU-MIMO with Gaussian signaling~\cite{WeiYuTIT2004}. When CSI is only available at the receiver, the Gaussian capacity of MU-MIMO was obtained with Gaussian signaling~\cite{tse2005fundamentals}. For binary input distributions, the constrained capacity of single-user MIMO (SU-MIMO) with correlated channel matrices was derived by the replica method~\cite{Muller2003TSP}. For arbitrary input distribution, the constrained capacity of SU-MIMO was obtained with IID channel matrices~\cite{Barbier2017arxiv, ReevesTIT2019}, or unitarily-invariant channel matrices~\cite{Barbier2018b, LeiOptOAMP}, by using the adaptive interpolation method and random matrix theory~\cite{Barbier2018b, Barbier2017arxiv, ReevesTIT2019}, or using the MMSE and decoupling properties of approximate message passing (AMP)-type algorithms~\cite{LeiTIT2021, LeiOptOAMP}. For code-division multiple access (CDMA) with arbitrary input distributions and correlated channel matrices, the constrained capacity was obtained by the replica method~\cite{muller2009replica}. 

At present, there is a lack of theoretical analysis on the constrained capacity of GMU-MIMO with arbitrarily fixed input distributions and unitarily-invariant channel matrices.

\subsection{Practical Capacity Optimal Receivers of MU-MIMO}
To achieve the capacity of MU-MIMO, a lot of literature focused on the designs of parallel interference cancellation (PIC) receivers\cite{XiaojunTIT2014, LeiTSP2019,YuhaoTWC2018}. With a properly designed forward error correction (FEC) code, Turbo receiver is Gaussian capacity approaching with Gaussian signaling~\cite{LeiTSP2019,YuhaoTWC2018,XiaojunTIT2014}. Since transmitters cannot acquire CSI, a well-designed single-user code with linear precoding and iterative linear minimum mean square error (LMMSE) detection of SU-MIMO~\cite{XiaojunTIT2014} cannot be applied to MU-MIMO without user collaboration. For MU-MIMO with Gaussian signaling and CSI only available at the receiver, the optimality of Turbo-LMMSE was proved to achieve the Gaussian sum capacity\cite{LeiTSP2019,YuhaoTWC2018}. Nevertheless, for non-Gaussian discrete signaling, these Turbo receivers\cite{XiaojunTIT2014, LeiTSP2019,YuhaoTWC2018} are not capacity optimal anymore.

To address the above issue, AMP with properly designed FEC codes can achieve the constrained capacity of SU-MIMO with IID channel matrices and arbitrary input signaling \cite{LeiTIT2021}. In addition, for non-Gaussian signaling, it was shown that the achievable rate of AMP is higher than that of Turbo-LMMSE. However, AMP is limited to IID channel matrices~\cite{MaTWC2019}. For non-IID channels, AMP performs poorly or even diverges~\cite{Vila2015ICASSP,manoel2014sparse,Rangan2017TIT}, such that the results in\cite{LeiTIT2021} are invalid for GMU-MIMO.

In summary, the existing results are either limited to Gaussian signaling \cite{LeiTSP2019,YuhaoTWC2018,XiaojunTIT2014} or IID channels~\cite{LeiTIT2021}. 
\vspace{-0.1cm}
\subsection{Motivation}

To overcome the restriction of AMP, orthogonal AMP (OAMP) \cite{MaAcess2017} and vector AMP (VAMP)\cite{Rangan2019TIT} were proposed to offer improved performance in a wide range of unitarily-invariant matrices. Due to the equivalence of OAMP and VAMP, they are referred to as OAMP/VAMP in this paper. For un-coded systems, since the mean square error (MSE) of OAMP/VAMP can reach the MMSE estimated by 
the replica method when the compression rate of the system is larger than a certain value, OAMP/VAMP is Bayes (MMSE) optimal\cite{MaAcess2017,Rangan2019TIT,Kabashima2006, Barbier2017arxiv}. However, Bayes (MMSE) optimality cannot generally guarantee error-free performance. Achievable rate is a key measurement for coded systems with error-free recovery. As a result, a coded system is capacity optimal when the maximum achievable rate is equal to the constrained capacity. The capacity optimality of OAMP was proved in coded SU-MIMO systems~\cite{LeiOptOAMP}.
 
At present, it is still a lack of rigorous analysis on the information-theoretical limits of OAMP/VAMP for GMU-MIMO. Another key challenge is to design a practical information-theoretical optimal receiver that can achieve the constrained sum capacity and meet the different users’ rate requirements, especially for a large number of users.

\subsection{Contributions}
To address the above challenges in GMU-MIMO, we obtain the constrained sum capacity inspired by the replica method and propose a capacity optimal transceiver for GMU-MIMO. It is complicated to design transceivers for a completely asymmetrical GMU-MIMO that all users may have different rates. Therefore, group asymmetry is developed to make a good tradeoff between implementation complexity and rate allocation. Specifically, based on the celebrated mutual information and minimum mean-square error (I-MMSE) lemma~\cite{GuoTIT2005} and the MMSE optimality of OAMP/VAMP~\cite{Barbier2018b,Kabashima2006}, the constrained sum capacity is calculated, which equals the area covered by the MMSE transfer curves of OAMP/VAMP. Then, a practical multi-user OAMP/VAMP (MU-OAMP/VAMP) receiver is proposed for GMU-MIMO. A design principle of multi-user codes is presented for MU-OAMP/VAMP to achieve the constrained capacity region of group asymmetric GMU-MIMO. Moreover, a kind of multi-user low-density parity-check (MU-LDPC) code is designed for MU-OAMP/VAMP.

The major contributions of this paper are summarized as follows.

\begin{enumerate}
	\item Group asymmetric GMU-MIMO is developed and the constrained sum capacity of GMU-MIMO is obtained.
	\item The achievable rate and capacity optimality of MU-OAMP/VAMP are analyzed and proved, based on which the design principle of multi-user codes is presented.
	\item A  kind  of  capacity-approaching  MU-LDPC  code  is  designed  for  MU-OAMP/VAMP. Numerical results show that the finite-length performances of the proposed MU-LDPC codes with MU-OAMP/VAMP are about 2 dB from the constrained sum capacity and outperform those of the existing state-of-art methods.
\end{enumerate}
\emph{Note:} Due to the limitation of pages, detailed Theorems proofs, MU-OAMP/VAMP analysis, and MU-LDPC coding optimizations are given in a full version\cite{chi2021capacity} of this paper.

\begin{figure}[!t]
	\centering
	\includegraphics[width=0.9\columnwidth]{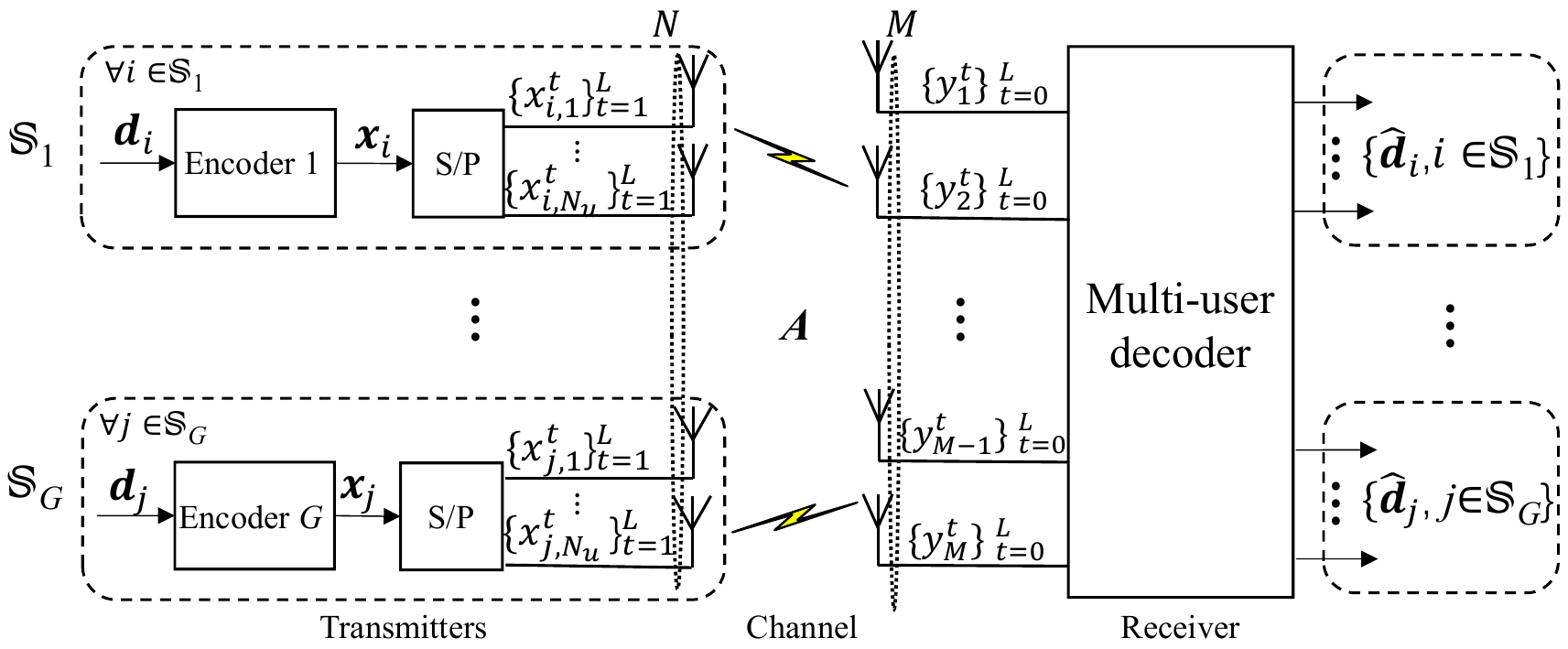}\vspace{-2mm}
	\caption{Illustration of a GMU-MIMO system with $K$ users partitioned into $G$ groups, and a receiver included a multi-user decoder. S/P denotes serial-to-parallel conversion. The number of transmitted and received antennas are $N$ and $M$ respectively.}\label{Fig:GMU-MIMO} \vspace{-2.5mm}
\end{figure}
\vspace{-1.5mm}
\section{System Model}
Fig.~\ref{Fig:GMU-MIMO} illustrates an uplink GMU-MIMO system with $K$ transmit users and one receiver. Total of $N$ transmit antennas are employed by the $K$ users and each user has $N_u=N/K$ antennas. The receiver has $M$ receive antennas. Users are equally partitioned into $G$ groups, where each group has $K/G$ users. Let  $\mathbb{S}_1, ...,\mathbb{S}_G$ be the sets that include the user indices of the $G$ groups, where $\mathbb{S}_i \cap \mathbb{S}_j = \emptyset$ for $i \ne j$, $i,j\in\mathbb{G}=\{1,...,G\}$, and $\mathbb{S}_1 \cup \mathbb{S}_2 \cdots \cup \mathbb{S}_{G} = \mathbb{K} \equiv \{1,\cdots, K\}$. Users in the same group employ the same encoder with the same code rate and that in different groups employ different encoders with different code rates. 

At the transmission side, since the processing of each user's data is similar, we describe the transmission of user  $i\in\mathbb{S}_1$.  Message vector $\bf{d}_i$  is encoded by encoder 1 and the output codeword is denoted by  $\bf{x}_i$. We assume $\bf{x}_i$ is a modulated signal vector whose entries are from a constellation set $\cal{S}$.  A serial-to-parallel conversion (S/P) is employed to produce the transmit signals over each antenna. Suppose the length of $\bf{x}_i$ is $N_uL$ for a given integer $L$. Codeword $\bf{x}_i$ is split into $N_u$ length-$L$ vectors $\bf{x}_{i,n}, n=1,...,N_u$, and $\bf{x}_{i,n}=\{x^t_{i,n}\}_{t=1}^L$ is transmitted  over antenna $n$. Here $L$ is the total transmission time for codeword $\bf{x}_i$. At time $t$, the transmittion signals by user $i$ over the $N_u$ antennas are written as $\bf{x}_i^t=(x^t_{i,1},...,x^t_{i,N_u})$. The all transmitted signals of $K$ users are denoted as $\bf{x}^t=(x^t_{1,1},..., x^t_{G,N_u}) \in \mathbb{C}^{N \times 1}$ which satisfy the power constraint $\tfrac{1}{N}\mr{E}\{\|\bf{x}^t\|^2\}=1$.

The receiver obtains signal $\bf{y}^t= [y_1^t, ..., y_M^t]^{\rm{T}}$, given by
\BE\label{Eqn:gmu_recv}
\bf{y}^t=\bf{A}\bf{x}^t+\bf{n}^t,\;\;  t=1,\dots,L,
\EE
where $\bf{A}\in \mathbb{C}^{M\times N}$ is a quasi-static fading channel matrix and $\bf{n}^t \sim\!\mathcal{CN}(\mathbf{0},\sigma^2\bm{I})$ is an AWGN vector. Note that $\bf{A}$ is right-unitarily-invariant and let SVD of $\bf{A}$ be $\bf{A} = \bf{U}\bf{\Lambda}\bf{V}$, where $\bf{U},\bf{V} \in \mathbb{C}^{M\times M}$ are unitary  matrices, and $\bf{\Lambda}$ is a rectangular diagonal matrix. $\bf{U}$, $\bf{V}$, and $\bf{\Lambda}$ are mutually independently, and $\bf{V}$ is Haar-distributed~\cite{MaAcess2017, MaTWC2019, Poor2021TWC}. Without loss of generality, we assume $\tfrac{1}{N}{\rm tr}\{\bf{A}^{\rm{H}}\bf{A}\}=1$ and the signal-to-noise ratio (SNR) is defined as  ${snr} = \sigma^{-2}$.
Based on $\bf{y}^t$, a multi-user decoder is employed to recover the $K$-user's messages.

\begin{figure}[!t]
	\centering
	\includegraphics[width=0.8\columnwidth]{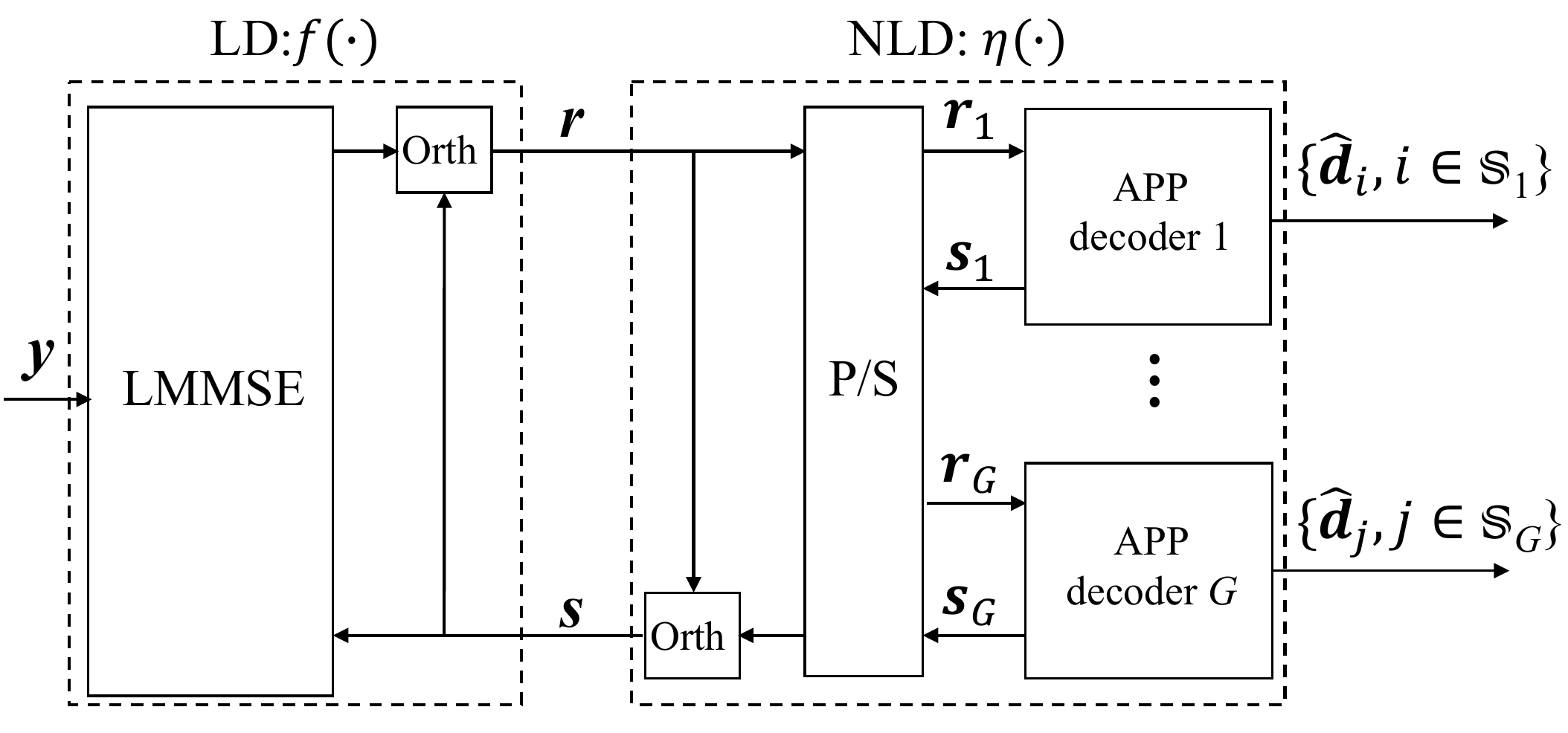} \\ \vspace{-2mm}
	\caption{MU-OAMP/VAMP receiver consists of an LD and an NLD, where the LD employs LMMSE detection and the NLD consists of $G$ APP decoders. ``Orth'' represents the orthogonalization of input and output  of LD and NLD.}\label{Fig:MU-OAMP}
\end{figure} 
\section{Multi-user OAMP/VAMP Receiver and Capacity Characterization of GMU-MIMO}
\subsection{MU-OAMP/VAMP Receiver}
\subsubsection{MU-OAMP/VAMP}
Since the detection process of \eqref{Eqn:gmu_recv} in each transmission time is the same, we omit the time index $t$ in the rest of this paper for simplicity.

As shown in Fig.~\ref{Fig:MU-OAMP}, MU-OAMP/VAMP consists of a linear detector (LD) and a non-linear detector (NLD), which employ LMMSE detection for linear constraint in~\eqref{Eqn:gmu_recv} and  \emph{a-posteriori probability}  (APP) decoding for code constraint $\bf{x} \in {\cal{C}}$, respectively. The detailed process is given as follows:
\BS\label{Eqn:MU-OAMP}\begin{align}
&\mathrm{LD:} \;\;\;\;\;\bf{r}\!=\! f(\bf{s}) \!=  c_{\cal{L}} f_{\mr{lmmse}}({\bf{s}}) +(1-c_{\cal{L}}) \bf{s},\label{Eqn:MULD}\\
&\mathrm{NLD:} \;\; \bf{s} = \eta(\bf{r})= c_{\cal{C}} \eta_{\mr{mmse}}({\bf{r}}) + (1-c_{\cal{C}}) \bf{r},\label{Eqn:MUNLD}
\end{align}
\ES
where $\bf{r}=[r_1, ..., r_N]^{\mr{T}}$ and $\bf{s}=[s_1, ..., s_N]^{\mr{T}}$ denote the LD and NLD estimations of $\bf{x}$ respectively, subscripts $\cal{L}$ and $\cal{C}$ indicate linear constraint and code constraint respectively. Let ${\bf{r}}_g$ and $\bf{s}_g$ be the input-output estimations of decoder~$g$ for the users in group $g$, $\forall g \in \mathbb{G}$.

In \eqref{Eqn:MU-OAMP}, the local estimation functions of LD and NLD are
\BS
\begin{align}
f_{\mr{lmmse}}({\bf{s}})&\equiv[{ snr }\bf{A}^{\rm{H}}\bf{A} +v_{s}^{-1}\bf{I}]^{-1}[{ snr }\bf{A}^{\rm{H}}\bf{y}+ v_{s}^{-1}{\bf{s}}],\label{Eqn:lmmse_Ax}\\
\eta_{\mr{mmse}}({\bf{r}})&\equiv \mr{E}\{ \bf{x} |\bf{r},\bf{x}\in {\cal{C}}\},\label{Eqn:mmse_Ck}
\end{align}
\ES
where $ snr $ denotes the given SNR. 
The parameters $c_{\mathcal{L}}$ and $c_{\mathcal{C}}$ in \eqref{Eqn:MU-OAMP} are given by
\begin{align}
c_{\cal{L}}=\frac{v_{{s}}}{v_{{s}}- \Omega_{\cal{L}}(v_{s}^{-1})}  \quad {\rm and} \quad c_{\cal{C}} = \frac{v_{r}}{v_{{r}}- \Omega_{\cal{C}}(\rho)}, 
\end{align} 
where  $v_s$ and $v_r$ are the variances of $\bf{s}$ and ${\bf{r}}$ respectively, and the MMSE (per transmit antenna) functions are defined as
\BS\begin{align}
\Omega_{\cal{L}}(\rho)  &\equiv \tfrac{1}{N}{\mr{Tr}}\{[{ snr }\bf{A}^{\rm{H}}\bf{A}+\rho\bf{I}]^{-1}\},\label{Eqn:MU_rlmmse}\\
\Omega_{\cal{C}}(\rho) & \equiv \tfrac{1}{G}\!\!\textstyle\sum\limits_{g=1}^{G}\Omega_{{\cal{C}}_g}(\rho), \label{Eqn:MU_rlmmseavg}\\ 
\Omega_{{\cal{C}}_g}(\rho) &\equiv \tfrac{G}{N} \mr{E}\big\{\|\bf{x}_g-\eta_{\mr{mmse}}( {\sqrt{\rho}} \bf{x}_g + {\bf{z}})\|^2\big\},\label{Eqn:MU_slmmse}
\end{align} \ES
where $\bf{x}_g$ is the signal vector of the users in $\mathbb{S}_g$, and ${\bf{z}}\sim \mathcal{CN}(0,\bf{I})$ is independent of $\bf{x}_g$. 

Note that the orthogonal operations~included in (\ref{Eqn:MULD}) and~(\ref{Eqn:MUNLD}) are necessary to make the input-output estimated errors of LD and NLD uncorrelated during the iteration detection process, which are denoted as ``Orth'' in Fig.~\ref{Fig:MU-OAMP}. It is proved that the orthogonalization is the key factor for exact state evolution of OAMP/VAMP~\cite{MaAcess2017,LeiOptOAMP}. 

\subsubsection{State Evolution}
LD and NLD are exactly characterized by the state evolution (SE), in which the transfer functions are consisted of SNR function $\psi_{\cal{C}}(\rho)$ and MSE function $\phi_{\cal{L}}(v)$ as follows.
\BS\label{Eqn:MUSE}
\begin{align}
&\!\!\!\mathrm{LD:}
\;\;\;\;\rho \equiv \phi_{\cal{L}}(\bar{v}) = \big[\tfrac{1}{N}\|\bf{r}-\bf{x}\|^2\big]^{-1},\label{Eqn:MUSE_LD}\\
&\!\!\!\mathrm{NLD:}\;\; \bar{v} \equiv \psi_{\cal{C}}(\rho) = \tfrac{1}{G}\textstyle\sum\limits_{g=1}^{G} \psi_{{\cal{C}}_g}(\rho) = \tfrac{1}{G}\sum\limits_{g=1}^{G}\|\bf{s}_g-\bf{x}_g\|^2.\label{Eqn:MUSE_NLD}
\end{align}
\ES

The following assumption, confirmed by numerical experiments in \cite{MaTWC2019}, shows the approximate IID Gaussianity of MU-OAMP/VAMP, which is critical to simplify the design and analysis of MU-OAMP/VAMP. 
\begin{assumption}[Approximate IID Gaussianity]\label{Lem:GA_MUSE} Let $M,N \to \infty$ with a fixed $\beta=N/M$ and $\bf{x} \in {\cal{C}}$.
	The input of NLD can be regarded as $\bf{r}  = \bf{x} + \rho^{-1/2}{\bf{z}}$ with ${\bf{z}}\sim \mathcal{CN}(\bf{0},\bf{I})$ independent of $\bf{x}$. As a result, (\ref{Eqn:MUSE_LD}) and (\ref{Eqn:MUSE_NLD}) are rewritten as 
	\BS\label{Eqn:MU-TF} \begin{align}
	\mathrm{LD:}& \quad   \rho =\phi_{\cal{L}}({\bar{v}})  = [{\Omega_{\cal{L}}({\bar{v}}^{-1})}]^{-1} -  {\bar{v}}^{-1},\label{Eqn:TFLD}\\
	\mathrm{NLD:}& \quad    \bar{v} = \psi_{\cal{C}}(\rho) = \left([\Omega_{\cal{C}}(\rho)]^{-1}- \rho \right)^{-1},\label{Eqn:TFNLD}
	\end{align}\ES
	where $\bar{v} \in[0,1]$ and $\phi_{\cal{L}}(0)=\tfrac{ snr }{N}{\rm tr}\{\bf{A}^{\rm{H}}\bf{A}\}={snr}$ due to the normalized singular values of $\bf{A}$, i.e., $\tfrac{1}{N}{\rm tr}\{\bf{A}^{\rm{H}}\bf{A}\}=1$.
\end{assumption}

Define $\phi_{\cal{L}}^{\mr{inv}}(\cdot)$ as the generalized inverse function of $\phi_{\cal{L}}(\cdot)$ and  $\varphi_{\cal{L}}(\rho) \equiv \big(\rho + [\phi_{\cal{L}}^{\mr{inv}}(\rho)]^{-1} \big)^{-1}$. As a result, for simplicity, we redefine $v$ as ${v}\equiv{\bar{v}}$, such that the transfer functions in \eqref{Eqn:MUSE} are transformed equivalently as follows
\BS\label{Eqn:NewMUSE}
\begin{align}
\mathrm{LD:} \;\;   & {{v}} = \varphi_{\cal{L}}(\rho), \label{Eqn:allocateV}\\
\mathrm{NLD:}  \;\; & {{v}} = \Omega_{\cal{C}}(\rho).
\end{align}\ES

\begin{figure}[!t]
	\centering
	\includegraphics[width=0.58\columnwidth]{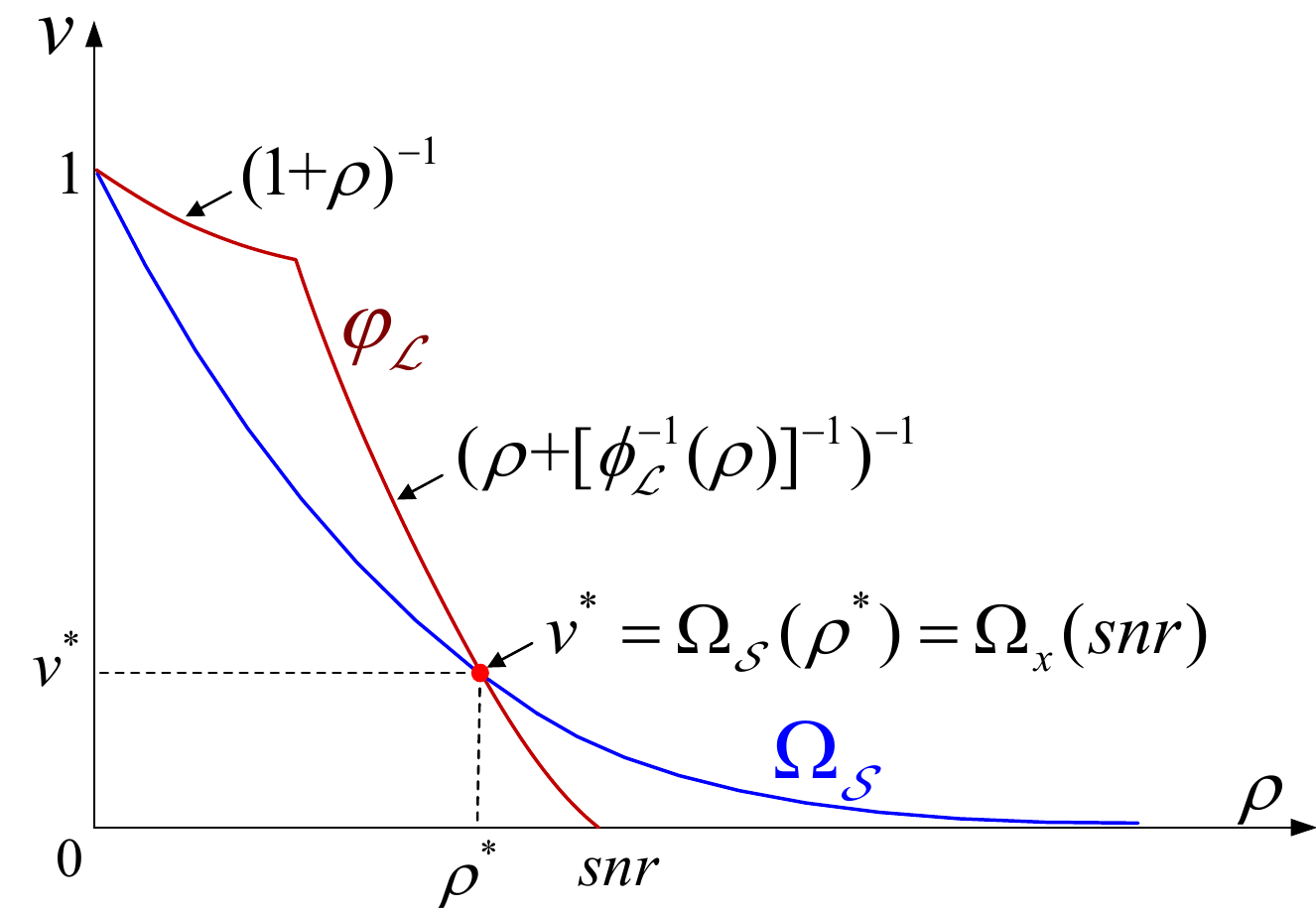}\\ 
	\caption{Illustration for transfer functions of MU-OAMP/VAMP in GMU-MIMO, where $\Omega_{\cal{S}}$ denotes the MMSE of constellation constraint and $(\rho^*,v^*)$ denotes the fixed point between $\Omega_{\cal{S}}$ and $\phi_{\cal{L}}$.}\label{Fig:un-coded_chart}
\end{figure}
\vspace{-0.5cm}
\subsection{Constrained Sum Capacity of GMU-MIMO} 
For simplicity of discussions, we define an un-coded GMU-MIMO as
\BE\label{Eqn:cons_gmu_recv}
\bf{y}=\sqrt{\rho}\bf{A}\bf{x}+\bf{z},
\EE
where $\bf{z}\sim \mathcal{CN}(\bf{0}, \bf{I})$ is a random noise vector, $\bf{x}=\{x_i\}$ with $x_i\sim P_{\mathcal{S}}(x_i), \forall i$, and $\mathcal{S}$ denotes a constellation.  For convenience, we define
\BS\label{Eqn:Omega_S}\begin{align} 
\Omega_{\mathcal{S}}(\rho) &\equiv \tfrac{1}{N} \text{mmse} \big(\bf{x}|\sqrt{\rho}\bf{x}+\bf{z}, x_i\sim P_{\mathcal{S}}(x_i), \forall i\big),\\
\Omega_{x}(\rho) &\equiv \tfrac{1}{N} \text{mmse} \big(\bf{x}|\sqrt{\rho}\bf{A}\bf{x}+\bf{z}, x_i\sim P_{\mathcal{S}}(x_i), \forall i\big), \\
\Omega_{{A}x}(\rho) &\equiv \tfrac{1}{N} \text{mmse} \big(\bf{Ax}|\sqrt{\rho}\bf{A}\bf{x}+\bf{z}, x_i\sim P_{\mathcal{S}}(x_i), \forall i\big).
\end{align}\ES 
Note that all the MMSE functions in this paper are defined on per transmit antenna.

Following the I-MMSE lemma~\cite{GuoTIT2005}, the average constrained capacity of GMU-MIMO per transmit antenna is calculated by 
\BE\label{Eqn:const_C}\vspace{-0.1cm}
\bar{C} = \tfrac{1}{N}I(\bf{x}; \sqrt{snr}\bf{A}\bf{x}+\bf{z}) =\int_0^{snr} \!\!\!\Omega_{{Ax}}(\rho)\:d\rho, 
\EE
which is reduced to $\int_0^{snr} \Omega_{{x}}(\rho)\:d\rho$ for single-user channels. 
Then, the constrained sum capacity is obtained as
\BE\label{Eqn:const_SC}
C_{\mr{GMU-MIMO}}^{\mr{sum}} = N \bar{C}.
\EE

Next, we derive the expression of $\Omega_{{Ax}}(\rho)$ in~\eqref{Eqn:const_C}, denoted by the replica MMSE, using the properties of MU-OAMP/VAMP.

As shown in Fig.~\ref{Fig:un-coded_chart}, the iterative process between LD $\phi_{\cal{L}}(\rho)$ and NLD  $\Omega_{\cal{S}}(\rho)$ converges to a unique fixed point $(\rho^*, v^*)$, where $v^*=\big([\Omega_{\mathcal{S}} (\rho^*)]^{-1} - \rho^*\big)^{-1}$ based on~\eqref{Eqn:TFNLD}. The curve $\varphi_{\cal{L}}(\rho)$ is an upper bound of $ \Omega_{\cal{S}}(\rho)$ for $0\le \rho \le \rho^*$. 

\begin{lemma}[Replica MMSE $\Omega_{Ax}(snr)$]\label{Lem:MMSE_Ax}
According to the fixed point $(\rho^*, v^*)$, $\Omega_{Ax}(snr)$ is given by
\BS\label{Eqn:mmse_Ax}
\begin{align}
\Omega_{Ax}({ snr })&= \big(1\!-\!v^{*^{-1}}\Omega_{\mathcal{L}}(v^{*^{-1}})\big)/{ snr }\!=\! \rho^*\Omega_{\mathcal{S}}(\rho^*)/{ snr }\\
&= [1\!-\!v^{*^{-1}}\Omega_{x}({{ snr }})]/{ snr }\!=\!\rho^*\Omega_{x}({{ snr }})/{ snr }.
\end{align}\ES
\end{lemma}

Based on Lemma~\ref{Lem:MMSE_Ax}, \eqref{Eqn:const_C} and \eqref{Eqn:const_SC}, the constrained sum capacity in \eqref{Eqn:cons_gmu_recv} is given in the following theorem.

\begin{theorem}[Constrained Sum Capacity]\label{Them:Constrain_sum}
Assume that $\phi_{\cal{L}}(\rho)$ and  $\Omega_{\cal{S}}(\rho)$ has a unique fixed point $(\rho^*, v^*)$. Then the constrained sum capacity of GMU-MIMO is given by\vspace{-0.2cm}
\BE\label{Eqn:const_SC_Express}
\!\!\!\!\!C_{\mr{GMU-MIMO}}^{\mr{sum}}\!= \!{\log \left| \bf{B}(v^*) \right|}+N \Big(\!\log  \Omega_{\mathcal{S}}(\rho^*)+\!\int_{0}^{\rho^*}\!\!\!\!{\Omega_{\mathcal{S}}(\rho) d\rho}\Big), 
\EE
\vspace{-0.2cm}
where $\bf{B}(v) = v^{-1}\bf{I} + { snr }\bf{A}^{\rm{H}}\bf{A}$. 
\end{theorem}

\begin{figure}[!t]
	\centering
	\includegraphics[width=0.5\columnwidth]{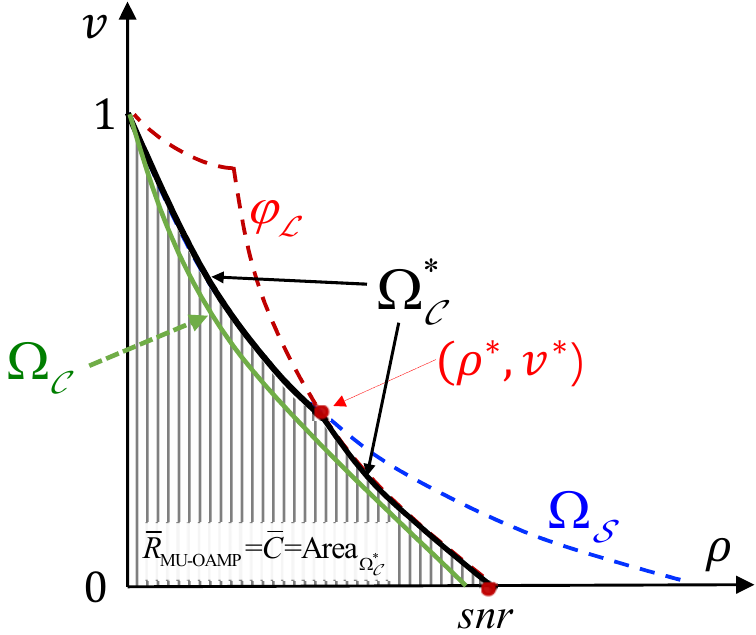}\\ \vspace{-0.2cm}
	\caption{Illustration of the relationship between rate and MMSE in MU-OAMP/VAMP. The minimum of $\varphi_{\cal L}$ in~\eqref{Eqn:NewMUSE} and $\Omega_{\cal{S}}$ in~\eqref{Eqn:Omega_S} is the upper bound for the optimal MMSE function $\Omega_{\cal{C}}^*$ in~\eqref{Eqn:MU_opt_NLD}. $\Omega_{\cal{C}}^*$ is the upper bound for $\Omega_{\cal{C}}$. The maximum achievable rate of OAMP/VAMP equals the average constrained capacity, which is the area covered by $\Omega_{\cal C}^*$ 
	}\label{Fig:SVTF_area}
\end{figure}

\section{Theoretical Capacity Optimality of MU-OAMP/VAMP}
In this section, we analyse the achievable rate of MU-OAMP/VAMP and give the design principle of multi-user codes.

\subsection{Capacity  Optimality of MU-OAMP/VAMP}
To verify the optimality of MU-OAMP/VAMP, we investigate the achievable rate of MU-OAMP/VAMP with error free performance.  Since locally optimal decoding should do better than symbol-by-symbol demodulation, $\Omega_{\mathcal{C}_g}(\rho)< \Omega_{\mathcal{S}}(\rho), \forall \rho \geq 0, \forall g.$ As a result, $\Omega_{\mathcal{C}}(\rho)< \Omega_{\mathcal{S}}(\rho),\;\; \forall \rho \geq 0.$

According to (\ref{Eqn:MU-TF}) and (\ref{Eqn:NewMUSE}), as shown in Fig.~\ref{Fig:SVTF_area}, MU-OAMP/VAMP receiver is error-free if and only if the NLD curve $\Omega_{\mathcal{C}}(\rho)$ lies below the LD $\varphi_{\mathcal{L}}(\rho)$. Therefore, the MMSE of a feasible coded NLD is defined as
\BE\label{Eqn:MU_opt_NLD}
\Omega_{\mathcal{C}}^*({\rho})= 
\min\{\Omega_{\mathcal{S}}(\rho),\; \varphi_{\mathcal{L}}(\rho)\},\;\;\;\forall\rho \in[0,  snr),  
\EE
and $\Omega_{\mathcal{C}}^*({\rho}) = 0$ for $\rho \ge  snr $.

Therefore, it is easily obtained as
\BE\label{Eqn:MU-upper_bound}
\Omega_{\cal{C}}({\rho}) < \Omega_{\mathcal{C}}^*({\rho}),\;\;\;\forall\rho \in[0,  snr).
\EE

Assume that there exist code-books $\{{\cal{C}}_g\}$ for $\{\mathbb{S}_g\}$, such that $\Omega_{\mathcal{C}}({\rho})$ can match $\Omega_{\mathcal{C}}^*({\rho})$, i.e., $\Omega_{\mathcal{C}}({\rho})=\tfrac{1}{G}\sum_{g=1}^{G}\Omega_{{\cal{C}}_g}(\rho) \to \Omega^*_{\cal{C}}({\rho})$. Thus, based on the I-MMSE lemma\cite{GuoTIT2005}, the achievable sum rate of MU-OAMP/VAMP is\vspace{-0.1cm}
\BE \label{Eqn:MU-A1}
R^{\rm{sum}}_{\rm{MU-OAMP}} \!\to\! \tfrac{N}{G}\!\textstyle\sum\limits_{g=1}^{G}\!\!R_{\mathcal{C}_g}  \!\!=\!N\!\!\int\limits_{0}^{snr} \!\!\Omega_{\cal{C}}^*({\rho}) d\rho \!=\! N\bar{R}_{\rm MU-OAMP}, \vspace{-0.1cm}
\EE
where $\bar{R}_{\rm MU-OAMP}\equiv\int_{0}^{snr} \Omega_{\mathcal{C}}^*({\rho}) d\rho$ denotes the average rate per transmit antenna.  

The following theorem verifies the capacity optimality of MU-OAMP/VAMP in GMU-MIMO.
\begin{theorem}[Sum Capacity Optimality]\label{Pro:MU_R_OAMP}
	Assume that $\Omega_{\mathcal{S}}(\rho)=\varphi_{\mathcal{L}}(\rho)$ has a unique positive solution $\rho^*$, and $v^*=\big([\Omega_{\mathcal{S}} (\rho^*)]^{-1} - \rho^*\big)^{-1}$. Then, $\bar{R}_{\rm MU-OAMP}=\bar{C}$ and
	\BE\label{Eqn:MU-achie_rate}
	R_{\rm{MU-OAMP/VAMP}}^{\rm{sum}} = C_{\mr{GMU-MIMO}}^{\mr{sum}}.
	\EE 
\end{theorem}

Fig.~\ref{Fig:SVTF_area} shows that the average constrained capacity of GMU-MIMO $\bar{C}$, the average rate of MU-OAMP/VAMP~${\bar{\it{R}}}_{\rm{MU-OAMP}}$, and the area covered by the optimal MMSE function of $\Omega_{\cal{C}}^*(\rho)$ are equal. 

\subsection{Multi-User Code Design for MU-OAMP/VAMP}\label{Sec:Cap_Opt}
\subsubsection{Constraints of Code Design} To reach the achievable rate of MU-OAMP/VAMP, $\{\Omega_{\mathcal{C}_g}(\cdot)\}$ in \eqref{Eqn:MU_slmmse} needs to be elaborately designed to make $\Omega_{\mathcal{C}}({\rho})$ in \eqref{Eqn:MU_rlmmseavg} match with $\Omega_{\mathcal{C}}^*({\rho})$ in \eqref{Eqn:MU_opt_NLD}. From~\eqref{Eqn:MU_opt_NLD}-\eqref{Eqn:MU-A1}, the constraints of $\{\Omega_{\mathcal{C}_g}(\cdot)\}$ are give in Property~\ref{Pro:Const_code}.
\begin{property}\label{Pro:Const_code}
	The MU-OAMP/VAMP is optimal when $\{\Omega_{\mathcal{C}_g}(\cdot)\}$ satisfy the following conditions.
	\begin{itemize}
		\item $\{\Omega_{\mathcal{C}_g}(\rho), \; \forall g\in \mathbb{G}\}$ are monotone decreasing in $\rho\geq0$.
		\item $\Omega_{\mathcal{C}}(\rho)=\tfrac{1}{G}\Sigma_{g=1}^{G}\Omega_{\mathcal{C}_g}(\rho)\to\Omega_{\mathcal{C}}^*(\rho)$.
		\item $0\leq\Omega_{\mathcal{C}_g}(\rho)<\Omega_{\mathcal{S}}(\rho), \;\: \forall g\in \mathbb{G}$.
	\end{itemize}
\end{property}

Based on Property~\ref{Pro:Const_code} and \eqref{Eqn:MU_opt_NLD}, we have \vspace{-0.2cm}
\BE\label{Eqn:Omega_n}
\Omega_{\mathcal{C}_g}(\rho)= \left\{ \begin{array}{l}
	\Omega_{\mathcal{S}}(\rho), \qquad\quad\; 0\leq\rho<\rho^* \vspace{-0.1cm}\\
	\zeta_g\big(\varphi_{\mathcal{L}}(\rho)\big), \quad\;\; \rho^*\leq\rho\leq { snr } \\ 
\end{array} \right.\!\!,  \;\;\;\forall g\in \mathbb{G},
\EE
and $ \Omega_{\mathcal{C}_g}(\rho)=0, \rho>{ snr }$. $\zeta_g(\cdot)$ is named as a variance allocation function, which  allocates the variance in \eqref{Eqn:allocateV} to user groups in different proportions.

\begin{lemma}[Sum Code Rate]\label{Lem:asym_sumrate}
	The sum rate of all users is 
	\BE\label{Eqn:asymRsum}
	\!\!\!R_{\mr{sum}}={\tfrac{N}{G}\textstyle\sum_{g=1}^{G}R_{\mathcal{C}_g}}=N\int_{0}^{snr}\!\!\!\Omega_{\mathcal{C}}^*({\rho}) d\rho=N\bar{R}_{\rm MU-OAMP/VAMP},
	\EE
	where $R_{\mathcal{C}_g}=\int_{0}^{\infty}\Omega_{{\cal{C}}_g}(\rho)d\rho$.
\end{lemma}

\subsubsection{Symmetric Systems} Consider $\{\zeta_g(x)=x, g\in \mathbb{G}\}$, which corresponds to the symmetric case that all the users have the same transfer function and rate, i.e.
\BE
\Omega_{\mathcal{C}_g}(\rho)=\Omega_{\mathcal{C}}^*(\rho), \quad \mr{and} \quad \bar{R}_{\mr{user}}= \bar{R}_{\rm MU-OAMP/VAMP}.
\EE

\subsubsection{Group Asymmetric Systems} In asymmetric systems,  the users in the same group  have the same rate but the users in different group  have different rates, which are determined by the variance allocation function $\{\zeta_g(\cdot)\}$. For simplicity, let $v=\Omega_{\mathcal{C}}^*(\rho)$ and $v_g=\Omega_{\mathcal{C}_g}(\rho)$. Meanwhile, we consider the following constraint of $v_g$ for $v \in[0, \Omega_\mathcal{S}(\rho^*)]$ similar as~\cite{LeiTSP2019}:
\BE\label{Eqn: g2}
\gamma_i (v_i^{-1} -  c^*)=\gamma_g (v_g^{-1}-c^*), \quad\forall i,g\in \mathbb{G},
\EE
where $c^*=[\Omega_\mathcal{S}(\rho^*)]^{-1}$ and $\gamma_i$, $\gamma_g$ $\in [0, \infty)$. Note that \eqref{Eqn: g2} can be written as  
\BE\label{Eqn:v_i}
v_i = [c^* +\gamma_i^{-1}{\gamma_g}(v_g^{-1}-c^*)]^{-1} = (b_{ig}v_g^{-1}+c_{ig})^{-1},
\EE
where $b_{ig}=\gamma_i^{-1}{\gamma_g}\geq0$ and $c_{ig}=(1-b_{ig})c^*$ are fixed. Since $v_g^{-1}-c^*>0$, $v_i$ is increasing with $\gamma_i$ while decreasing with $\{\gamma_g, k\neq i\}$. Thus, we obtain the variance allocation function:
\BE
\zeta_g(v_g) \equiv v=\tfrac{1}{G}\textstyle\sum_{g=1}^G (b_{ig}v_g^{-1}+c_{ig})^{-1},
\EE
which is a monotone increasing function for $v_g$. Based on~\eqref{Eqn:v_i}, $v_g=\zeta_g^{-1}(v)$
which may not have close form but it can be obtained numerically.  In addition, $\zeta_g(\cdot)$ is increasing with $\gamma_g$ and decreasing with $\{\gamma_i, i\neq g\}$, such that the code rate $\{R_{\mathcal{C}_g}\}$  can be changed flexibly by adjusting~$\{\gamma_g\}$, which is useful in the practical multi-user codes design.

\subsubsection{Transfer Curves Adjustment} 
Following Property~\ref{Pro:Const_code},  $\{v_g\}$ should satisfy  
\BS\label{Eqn:cons_vk}
\begin{align}\label{Eqn:cons_vk1}
&v_g=\mr{min}\{\zeta_g(v),\Omega_{\cal{S}}(\rho)\}, \quad \forall g\in \mathbb{G},\\ \label{Eqn:cons_vk2}
&\tfrac{1}{G}\!\!\textstyle\sum\limits_{g=1}^G v_g\!=\!\frac{1}{G}\textstyle\sum\limits_{i=1}^G (b_{ig}v_k^{-1}\!+c_{ig})^{-1} \!=\! v,\;\; \forall v\!\in\! (\Omega_{\mathcal{S}}(\rho^*), 1].
\end{align}
\ES
\section{Numerical Results}
This section provides BER simulations and comparisons of the proposed MU-LDPC codes with MU-OAMP/VAMP in GMU-MIMO. The details of MU-LDPC code optimization for MU-OAMP/VAMP can be found in~\cite{chi2021capacity}.

We consider BER simulations for two user group GMU-MIMO. Each group has $250$ transmitted antennas. The total number of transmitted antenna $N=500$. The coding length of each user in each group is $10^5$, each user employ optimized MU-LDPC codes~\cite{chi2021capacity}, QPSK modulation is employed, and Channel matrix $\bf{A}$ is used $400$ times for the whole transmission, where  $N=500$, $N=333$, channel load $\beta=N/M=1.5$, and condition number $\kappa=50$. Note that the user rate $R_{{\cal{C}}_g}$ of each group is equal to $N_uR_{\rm{LDPC}}\log_2|{\cal{S}}_{\rm{QPSK}}|$.

\begin{figure}[!t]
	\centering
	\includegraphics[width=0.88\columnwidth]{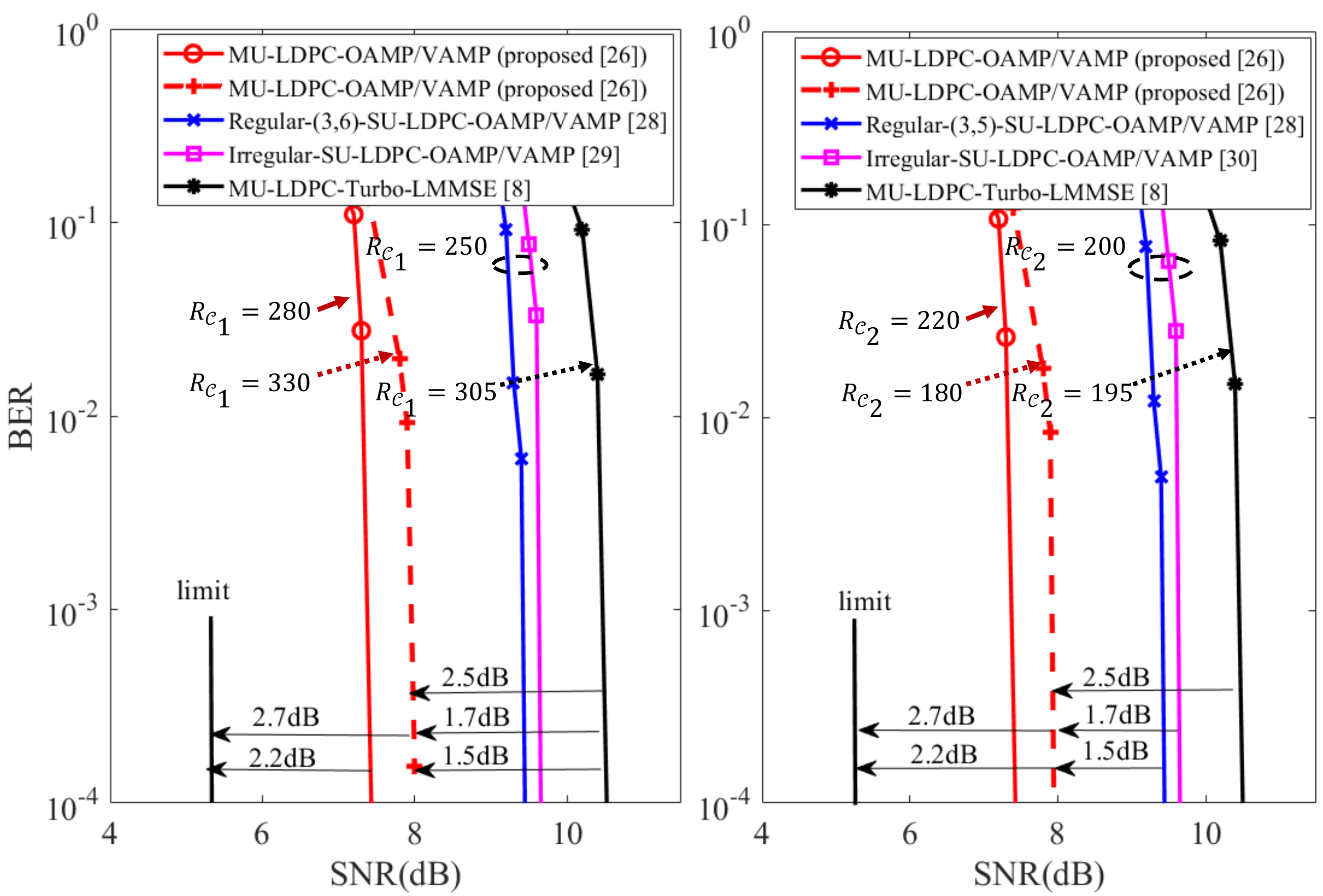}\\ 
	\caption{BER performances of the proposed MU-LDPC codes with $\{R_{\mathcal{C}_1}=330, R_{\mathcal{C}_2}=180\}$ and $\{R_{\mathcal{C}_1}=280, R_{\mathcal{C}_2}=220\}$ for MU-OAMP/VAMP, the regular (3,6), (3,5) SU-LDPC codes with $R_{\mathcal{C}_1}=250$ and $R_{\mathcal{C}_2}=200$~\cite{ryan2009channel} for MU-OAMP/VAMP, the well-designed irregular SU-LDPC codes with $R_{\mathcal{C}_1}=250$~\cite{Richardson2001} and $R_{\mathcal{C}_2}=200$~\cite{Kim2009} for MU-OAMP/VAMP, and the well-designed MU-LDPC codes with $R_{\mathcal{C}_1}=305$ and $R_{\mathcal{C}_2}=195$ for Turbo-LMMSE~\cite{LeiTSP2019}.}\label{Fig:ber_maxpoint_c} 
\end{figure}

Fig.~\ref{Fig:ber_maxpoint_c} shows that the proposed MU-LDPC codes with $\{R_{\mathcal{C}_1}=330, R_{\mathcal{C}_2}=180\}$ have about $1.7$~dB and $1.5$~dB performance gains over the regular (3,5), (3,6) SU-LDPC codes~\cite{ryan2009channel} and the well-designed irregular SU-LDPC codes~\cite{Richardson2001,Kim2009} with MU-OAMP/VAMP. This indicates that performances of Bayes-optimal MU-OAMP/VAMP with well-designed SU-LDPC codes are rigorously sub-optimal. Compared with the well-designed MU-LDPC codes with Turbo-LMMSE~\cite{LeiTSP2019}, the proposed MU-LDPC codes have about $2.5$~dB performances gains and higher transmission rates. These simulations verify that advantages of the proposed transceiver in GMU-MIMO.

\section{Conclusion}
This paper focuses on coded GMU-MIMO systems. The constrained sum capacity of GMU-MIMO is accurately calculated by using I-MMSE lemma and the Bayes optimal MU-OAMP/VAMP. The capacity optimal MU-OAMP/VAMP receiver with a matched error correction coding is considered to obtain an error-free recovery performance for GMU-MIMO. Based on the theoretical analysis of MU-OAMP/VAMP, the principle of multi-user code design is presented. Moreover, a kind of MU-LDPC code is optimized for GMU-MIMO and achieves better performances than the existing state-of-art methods.

\clearpage
\bibliographystyle{IEEEtran}
\bibliography{manuscript}

\end{document}